# Optimization of the magnetic properties of FePd alloys by severe plastic deformation

A. Chbihi[1], X. Sauvage[1*], C. Genevois[1], D. Blavette[1,2], D. Gunderov[3], A.G. Popov[4]

1- University of Rouen, CNRS UMR 6634, Groupe de Physique des Matériaux, Faculté des Sciences, BP12, 76801 Saint-Etienne du Rouvray, France
2- Institut Universitaire de France, 103 bd Saint-Michel, 75005 Paris, France.
3- Institute for Physics of Advanced Materials, Ufa State Aviation Technical University, K. Marx 12, Ufa 450000, Russia
4- Institute of Metal Physics, 18 Kovalevskaya str., Ekaterinburg, 620041, Russia

* corresponding author : xavier.sauvage@univ-rouen.fr

**Abstract**

A FePd alloy was nanostructured by severe plastic deformation following two different routes: ordered and disordered states were processed by high pressure torsion (HPT). A grain size in a range of 50 to 150 nm is obtained in both cases. Severe plastic deformation induces some significant disordering of the long range ordered $L1_0$ phase. However, Transmission Electron Microscopy (TEM) data clearly show that few ordered nanocrystals remain in the deformed state. The deformed materials were annealed to achieve nanostructured long range ordered alloys. The transformation proceeds via a first order transition characterized by the nucleation of numerous ordered domains along grain boundaries. The influence of the annealing conditions (temperature and time) on the coercivity was studied for both routes. It is demonstrated that starting with the disorder state prior to HPT and annealing at low temperature (400°C) leads to the highest coercivity (about 1.8 kOe).

## 1. Introduction

Permanent magnets play a key role in various technologies like energy conversion, data storage, sensors and actuators. Hard magnetic alloys exhibit a concomitant broad magnetic hysteresis and high coercivity $H_c$ that usually result from shape anisotropy of single domain particles and/or magnetocrystalline anisotropy. Largest coercivities are reached for nanostructured alloys with high intrinsic magnetocrystalline anisotropy (like in alloy containing rare earth (RE) elements). Small grain sizes (below 100 nm) give rise to high coercivity but a random grain orientation does not lead to high remanences. However, if the grain size is below or close to 10 nm, exchange coupling between moments of neighbouring crystallite can significantly increase the remanence. [1] During the last decade it has been shown that severe plastic deformation (SPD) techniques could be efficiently used to design nanoscaled structures. [2, 3] Some successful attempts to optimize the properties of hard magnetic alloys were reported in literature: increase of the coercivity of a Cu-Co alloy processed by equal channel angular pressing (ECAP), [4] improvement of the toughness properties of a Fe-Cr-Co alloy processed by high pressure torsion (HPT) [5] and significant increase of the coercivity of RE-Fe-B alloys processed by ECAP or HPT. [6-9]

Long range ordered intermetallics obtained based on Fe or Co and noble metals like Pt or Pd exhibit high uniaxial magnetocrystalline anisotropy with potential large energy products, good mechanical [17] and corrosion properties. [11] They are attractive for both thin film device applications (high density magnetic recording media) [12] and permanent magnets. [13] In spite of their relatively high cost, these materials are attractive for some specific applications like medical implants or extreme conditions where high strength and excellent corrosion resistance are required. It has been shown that their coercivity strongly depends on grain size. [13, 14] This is related to grain boundaries that act as pinning sites for domain walls. Ultrafine grain materials are commonly achieved by SPD processes, [2, 3] and it has been demonstrated that grain sizes smaller than 100 nm can be obtained in intermetallics. [15, 16] The present study aims at using such processes to optimize the magnetic properties of an equi-atomic intermetallic FePd alloy by nanostructuring.

Intermetallic FePd alloys near the equi-atomic composition undergo a first order type order-disorder transformation at $T_c \approx 650°C$. [18] The ordered phase (L1$_0$) exhibits a significant tetragonality along the c-axis leading to a high magnetic anisotropy. Ordered FePd alloys are usually characterized by a high density of twins that relax strains resulting from the nucleation and growth of the tetragonal phase. Alternative thermo-mechanical processing routes have been designed to improve the magnetic properties. [19] Besides, extremely high values of $Hc$ (up to 2.5 kOe) have been reported for rapidly solidified ribbons and balled milled powder. [20] A record coercivity of about 3 kOe was reported by Yermakov and co-authors for aerosol FePd powders. [22] Thus nanostructuring by SPD seems a very attractive way to achieve a unique combination of magnetic, corrosion and strength for these alloys.

Two different routes have been explored in the present study: i) the ordered phase was directly processed by SPD to create small ordered domains with high angle grain boundaries, ii) the disordered alloy (quenched from the high temperature state) was processed by SPD followed by aging to nucleate and grow ordered nanoscaled grains.

## 2. Experimental

A Fe50-Pd50 alloy with an equi-atomic composition was cast from high purity Fe (99.5%) and Pd (99.9%) alloys. The as-cast material was homogenized at 950°C during 6h in Ar atmosphere and subsequently quenched in ice brine to freeze the disordered fcc phase. Two different states were processed by HPT: the as-quenched material (disordered) and the fully ordered alloy. The ordered state was obtained by annealing at 550°C during 15h. We can therefore define the two following routes:

(i) Route 1: nanostructuring by HPT of the disordered state followed by annealing to nucleate and grow nanoscaled ordered domains

(ii) Route 2: nanostructuring the fully ordered state to get directly a nanostructured ordered structure.

Materials were processed by HPT at room temperature under a pressure of 6 GPa. HPT samples (8 mm in diameter discs with a thickness of 0.2 mm) were subjected to different number of revolutions $N = 1, 5, 7$ and $10$.

Microstructures were characterized by Transmission Electron Microscopy (TEM). Observations were performed with a JEOL 2000FX and a JEOL 2100 microscopes operating at 200 kV. TEM samples were punched out of the HPT discs at a distance of 2.5 mm from the center (corresponding to a shear strain of about $\gamma \approx 400$ for 5 turns) and then electropolished (electrolyte: 82% acetic acid + 9% phosphoric acid + 9% ethanol (vol. %); temperature: 0°C; voltage: 30V).

The ordering kinetics was also characterized by X-ray diffraction (XRD). Spectra were recorded with a Brucker D8 system in Bragg-Brentano $\theta$-$2\theta$ geometry. The X-ray generator was equipped with a Co anticathode, using Co (K$\alpha$) radiation ($\lambda = 0.17909$ nm). The measurements were performed so that the out-of-plane component was the torsion axis of the sample processed by HPT. The original structures of the alloy after quenching (disordered state) and after ordering (ordered state) were analysed by XRD (data not shown here). The L1$_0$ long range ordered state is characterized by strong superstructure peaks ((001), (110), (201), (112)) and a split of ((200) and (202)) peaks resulting from the tetragonality of the lattice, [12] whereas these features are not observed for the fcc disordered state.

The evolution of the *Hc* was monitored using a VSM producing a maximum field of 6 kOe at room temperature.

To quantify the evolution of the mechanical properties as a function of the applied strain, microhardness measurements were performed on a Future-Tech 7E micromet with a load of 300 g.

## 3. Results and discussions

3.1) as deformed structures

*Disordered state processed by HPT*

The hardness of the disordered state significantly increases after HPT processing. In the early stage of the deformation, it increases only at the periphery of the HPT disc where the deformation is larger. Then, a saturation of the hardness was observed and for a number of turns larger than seven no more significant hardness gradient across the HPT disc diameter was found. After 7 revolutions the average hardness of the sample has increased from 175 ± 6 HV (kg/mm$^2$) (original hardness of disordered state) to 420 HV ± 20 HV (kg/mm$^2$).

The observation of the microstructure by TEM suggests that this dramatic increase of the hardness might be attributed to the formation of a nanoscaled structure by HPT (Fig. 1). Grain boundaries are hardly visible on the bright field image (Fig. 1(a)). However, the selected area electron diffraction pattern (SAED, Fig. 1 (b)) clearly exhibits the typical Debye-Scherrer rings of polycrystalline structures with very small crystallite size. The ring corresponding to the (111) fcc reflection was partly selected using an aperture to image in the dark field mode isolated grains (Fig. 1 (c)). The grain size distribution is large: most of grains are smaller than 50 nm, but there is also a significant amount of bigger grains with a size in a range of 100 to 150 nm.

*Ordered state processed by HPT*

As expected, the long range ordered $L1_0$ structure exhibits a higher hardness compared to the disordered state: $300 \pm 6$ HV (kg/mm$^2$) versus $175 \pm 6$ HV (kg/mm$^2$). It also increases after HPT processing but in a smaller amount compared to the disordered state (up to $405 \pm 15$ HV (kg/mm$^2$)). It is worth mentioning that a homogeneous hardness of the sample across the HPT disc diameter was observed for only five revolutions instead of seven for the disordered state. Hence, surprisingly, after HPT, the hardness of the ordered and of the disordered state are almost the same (respectively $420 \pm 20$ and $405 \pm 15$ HV (kg/mm$^2$)).

These two states processed by HPT were therefore analysed by XRD (data not shown here). Surprisingly the recorded spectra did not exhibit any significant difference. A typical disordered fcc structure without any superstructure peak or peak splitting was observed for both states. In agreement with some earlier studies on long range ordered intermetallics processed by SPD, [15, 16, 21] the ordered $L1_0$ phase was disordered by SPD.

TEM observations revealed a nanoscaled structure with a grain size distribution similar to that of the material processed by route 1 (Fig. 2). On the SAED pattern (Fig. 2 (b)), Debbye-Scherrer rings corresponding to the fcc disordered phase are clearly exhibited, confirming that the long range ordered $L1_0$ structure was disordered by HPT. However, a careful observation shows some superlattice reflections corresponding to few isolated long range ordered grains (indicated by arrows). Unfortunately, these reflections were not strong enough to allow a successful imaging of these ordered grains in the dark field mode. Although these data indicate that the ordered $L1_0$ phase was not fully disordered during the HPT process and few ordered crystallites remain, it is clear that route 2 (i.e. HPT of the ordered state) cannot directly produce a nanostructured ordered state with a potential attractive combination of high strength, good corrosion properties and high coercivity (a coercivity as low as about 200 Oe was indeed measured).

3.2) ordering of the nanostructures obtained by HPT

Since route 2 failed to produce directly nanostructured ordered domains because of SPD induced disordering, both states processed by HPT were annealed at different temperature to nucleate the low temperature equilibrium $L1_0$ ordered phase. The ordering kinetics and especially the competition between ordering and recrystallisation were mostly studied using TEM. Indeed, it was quite difficult to extract accurate information from XRD because of the small grain size and of lattice defects resulting from SPD that induce a significant peak broadening.

Fig. 3 shows the microstructure of the alloy processed through route 1 (aged 3h at 500°C for ordering). Although the mean grain size is significantly larger than that of as-deformed materials, it is still in the ultrafine range (50 to 200 nm). On the bright field image (Fig. 3(a)), a mixture of recrystallized grains (with a low defect density) and unrecrystallized grains (with strong distortion contrasts typical of crystalline defects) appears. The SAED pattern (set in Fig. 3(a)) clearly exhibits some superlattice reflections of the $L1_0$ ordered phase, indicating that some ordering occurred as expected. The main question, already raised by R.A. Buckley 30 years ago for the FeCo system, [23, 24] is the following: how do the recrystallization and the ordering process in a concomitant way to reduce the energy of the system? Although the ordering process needs in principle only a diffusion of atoms over very short distances (typically only one atomic distance), the ordering process in the equi-atomic FePd alloy, like in the FeCo alloy, is a first order transition controlled by the nucleation and growth of ordered domains. As shown on the dark field image (Fig. 3(b)), these ordered domains nucleate along grain boundaries and they grow in un-recrystallized nanograins leading progressively to the full transformation of the alloy.

The nucleation of these numerous ordered nanoscaled domains significantly affects the magnetic properties. The influence of the ordering temperature and of the processing route on the coercivity is plotted in the Fig. 4. In the early stage of ordering, the slope of the curve is steeper for higher annealing temperatures. This could be attributed to a faster ordering kinetic promoted by a higher atomic mobility. After longer annealing times, the coercivity saturates and then decreases as a result of the coarsening of ordered domains driven by the reduction of grain boundary areas. The maximum of coercivity is reached for the lowest ordering temperature at 400°C after about 40h of annealing. This maximum, close to 1.8 kOe, is much higher than those typically reported in the literature for bulk samples processed by other thermo-mechanical treatments (about 1.3 kOe) [13, 14]. This is a direct consequence of the SPD induced nanostructuring: the proportion of grain boundaries and thus of nucleation sites for ordered domains is much higher leading to a smaller ordered grain size at the end of the ordering process and thus to a higher coercivity. [13, 14] When the ordering temperature is increased, a lower coercivity is achieved because the ordering driving force is lower and the boundary migration faster, leading to a larger mean grain size after ordering. It is also interesting to note that route 2 leads to a significantly lower coercivity (20% less for an ordering temperature of 450°C, see Fig. 4).This could be attributed to the nanoscaled ordered domains that still exist after the HPT processing (Fig. 2). Indeed, most of these grains are probably bigger than the critical nuclei size for the ordering transformation. Thus, they will grow immediately once the ordering treatment would have start and at the end the material would contain a significant proportion of large ordered domains.

In spite of the significant enhancement of the magnetic properties of the FePd hard magnets achieved by HPT, they are unlikely to compete with CoPt or FePt alloys. However, considering the lowest cost of Pd comparing to Pt, nanostructured FePd alloys are promising, especially for applications in medical implants or in extreme conditions were an excellent combination of mechanical strength, corrosion resistance and high coercivity is required. Based on the processing route described in the present paper, it is believed that further refinement of ordered domains (down to 50-100nm) can be achieved leading to a coercivity probably up to 3 kOe. To reach this target, one may speculate that HPT processing at lower temperature (i.e. below room temperature) may give rise to a smaller grain size and in turn to better magnetic properties. Besides, processing the ordered state at higher temperature (between room temperature and the critical temperature, i.e. still in the ordered domain of the phase diagram) may prevent the SPD induced disordering which could directly give rise to a nanostructured ordered FePd alloy. These two options will be soon assessed by the present authors.

## 4. Conclusions

i) A similar grain size in a range of 50 to 150 nm was obtained for both the ordered and the disordered states processed by HPT.

ii) The ordered state is significantly disordered by HPT but TEM data show that few ordered nanocrystals remain.

iii) During annealing of the HPT processed alloys, ordered domains nucleate along grain boundaries and grow to fully transform un-recrystallised grains. The mean grain size increases during annealing but it remains in the ultrafine range (50 to 200 nm).

iv) A higher coercivity is achieved if the material prior to HPT is in the disordered state.

v) Annealing at low temperature (400°C) gives rise to the highest coercivity: about 1.8 kOe, which is significantly higher than that usually obtained by standard thermo-mechanical treatment of this alloys.

vi) It is believed that improvement of the SPD processing route could give rise to further refinement of ordered domains and a much higher coercivity. Such opportunity would make FePd alloys processed by HPT very attractive for some specific applications where an excellent combination of mechanical strength, corrosion resistance and high coercivity is required.

**Acknowledgements**

Dr R. Ravel-Chapuis from JEOL Ltd is gratefully acknowledged for providing access to the JEOL 2100 TEM and the image acquisition. The Russian authors express their gratitude to the Russian Foundation for Basic Research for the financial support for this work (projects No. 07-02-92180).

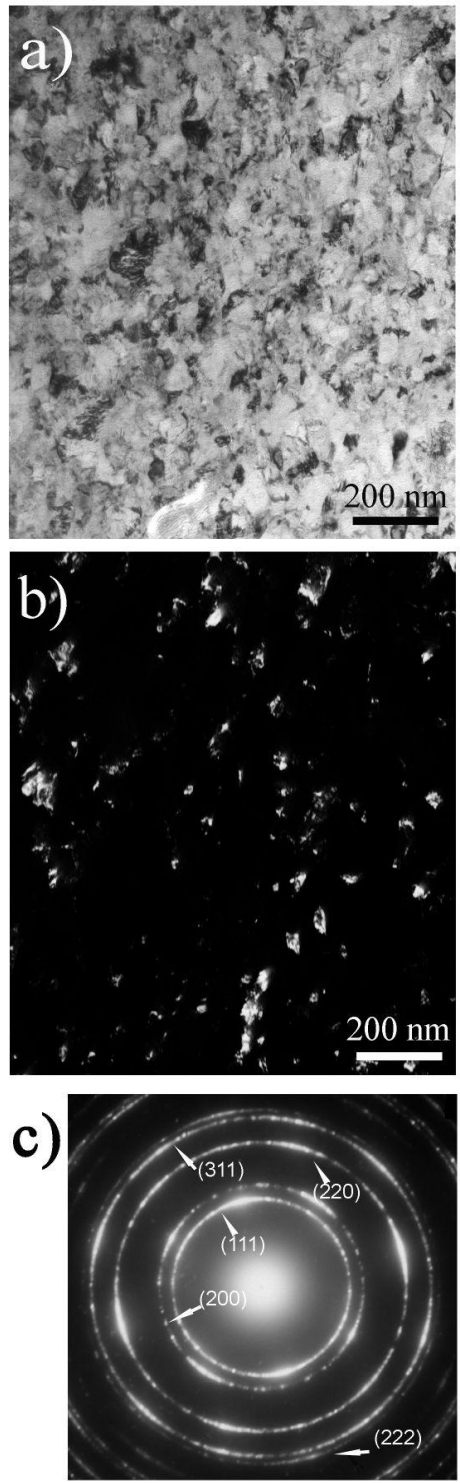

**Figure 1:** TEM images of the FePd alloy processed by HPT (route 1, 5 turns); Bright field image showing the nanoscaled structure (a), dark field image obtained by selecting with an aperture a part of the (111) fcc ring and showing few isolated nanoscaled grains (b), SAED pattern with Debye-Scherrer rings characteristic of a polycrystalline structure with a very small crystallite size (c).

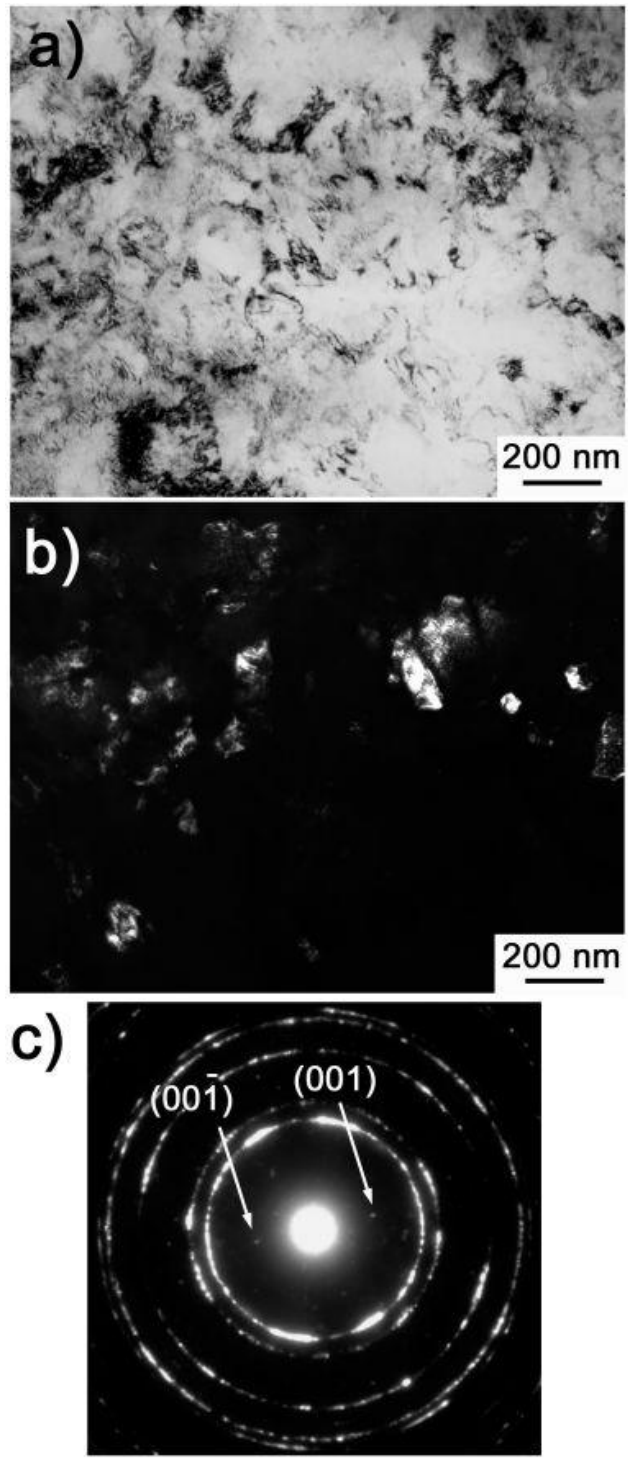

**Figure 2:** TEM images of the FePd alloy processed by HPT (route 2, 5 turns); Bright field image showing the nanoscaled structure (a), dark field image obtained by selecting with an aperture a part of the (111) fcc ring and showing few isolated nanoscaled grains (b), SAED pattern showing (001) superlattice reflections of the $L1_0$ ordered phase (arrowed) and indicating that few ordered grains remain.

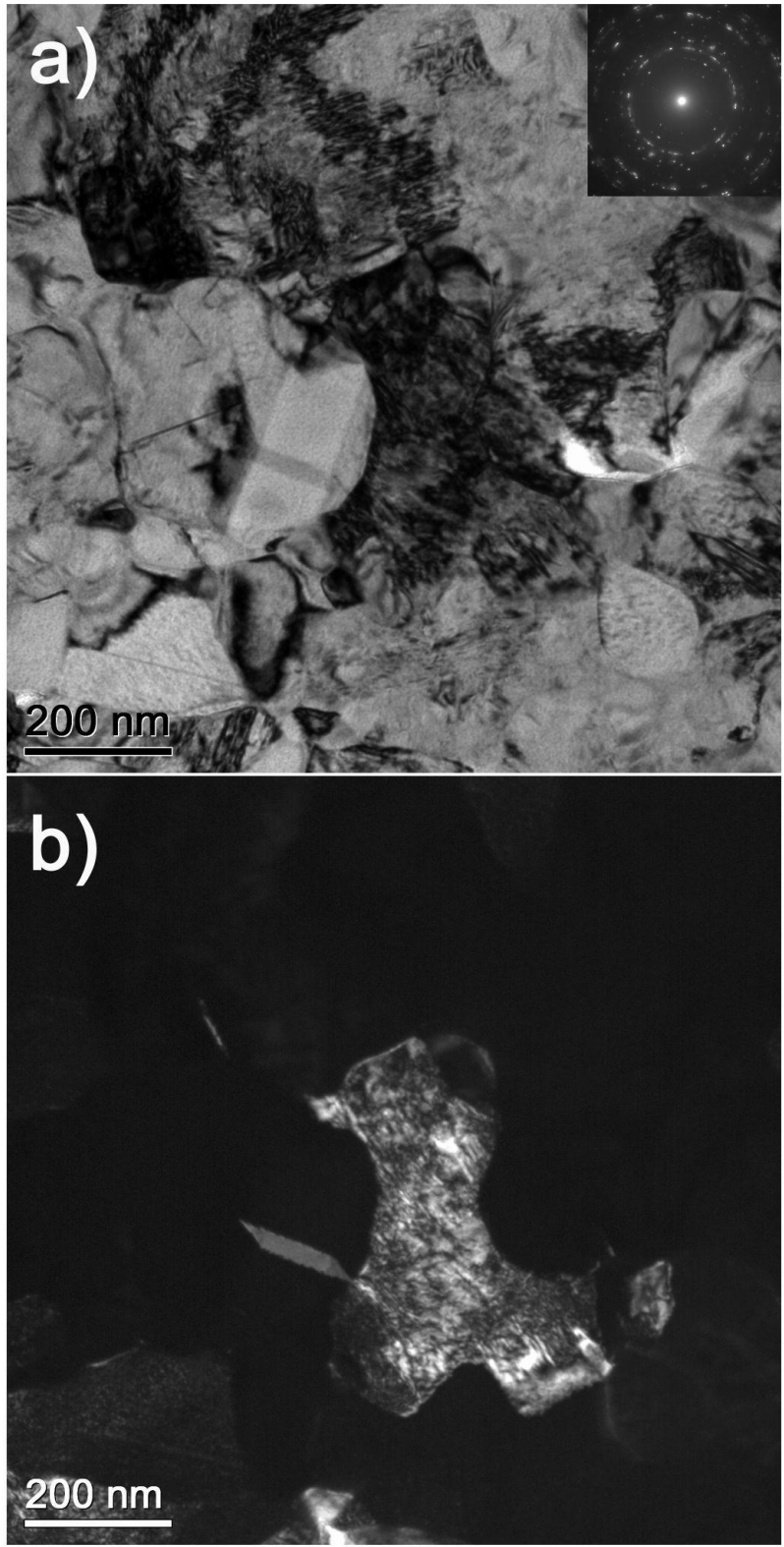

**Figure 3:** TEM images of the FePd alloy processed by HPT (route 1, 5 turns) and aged during 3h at 500°C; Bright field image showing the nanoscaled structure and SAED pattern (inset) showing superlattice reflections of the ordered $L1_0$ phase (a), dark field image showing a non recrystallised grain.

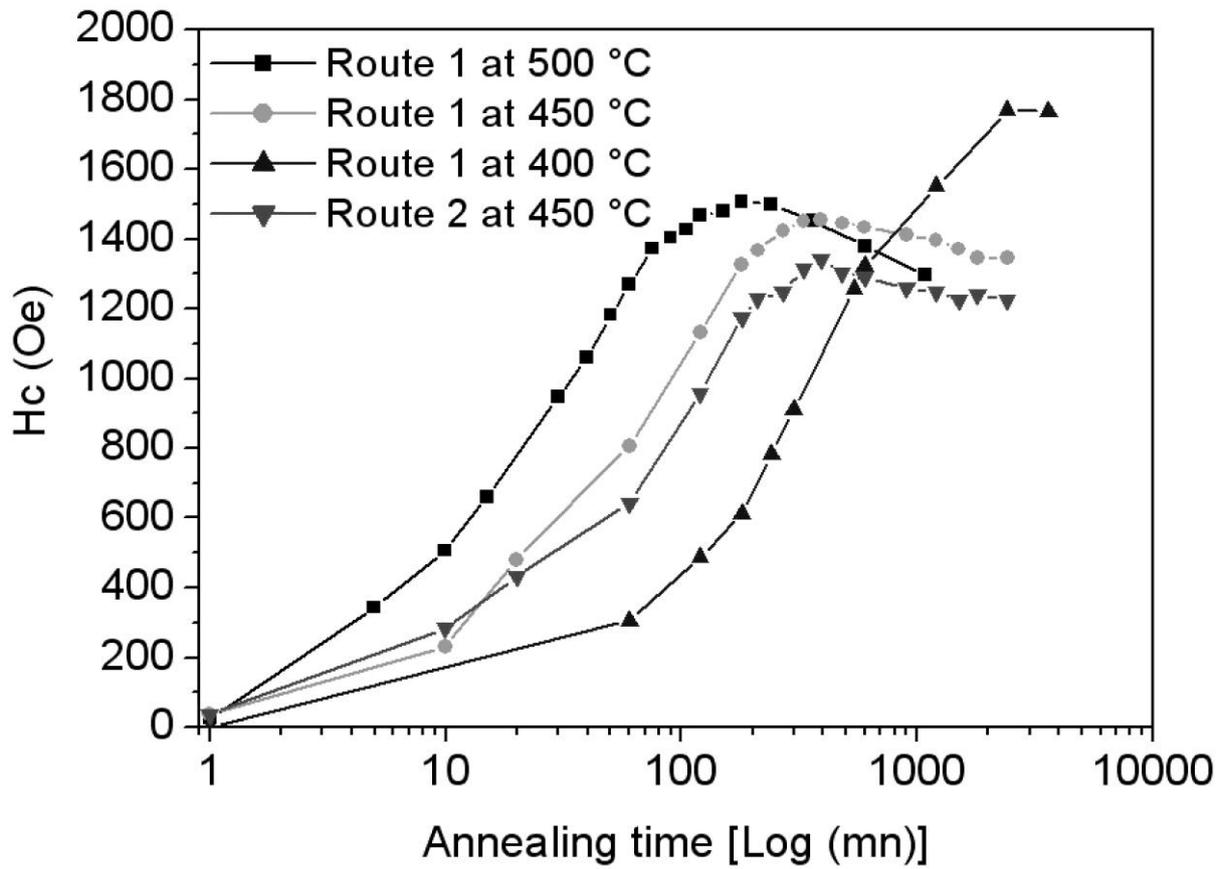

**Figure 4:** Evolution of the coercivity of the FePd alloy processed by route 1 and 2 as a function of the aging time for different temperatures. (Measurements performed on the outer part of the HPT disc).